\documentclass[final]{ias2}

\usepackage{graphicx} 
\usepackage{multirow}
\usepackage{array} 

\usepackage{hyperref} 

\begin{document}

\markboth{Pramana class file for \LaTeX 2e}{Debasish Das, et. al.}

\title{Quarkonia production at forward rapidity in Pb+Pb collisions at 
$\bf \sqrt{s_{\rm NN}}=2.76$~TeV with the ALICE detector}

\author[das]{Debasish Das}
\email{debasish.das@saha.ac.in; debasish.das@cern.ch}

\address[das]{Saha Institute of Nuclear Physics, 1/AF, Bidhan Nagar,
Kolkata 700064, India.\\
(For the ALICE Collaboration)}

\begin{abstract}
The study of formation of heavy quarkonia in relativistic heavy ion collisions 
provides important insight into the properties of the 
produced high density QCD medium. 
Lattice QCD studies show sequential suppression of quarkonia states with 
increasing temperature; which affirms that a full spectroscopy, can provide 
us a thermometer for the matter produced under extreme conditions in 
relativistic heavy ion collisions and one of the most direct probes 
of de-confinement. Muons from the decay of charmonium resonances are 
detected in ALICE Experiment in p+p and Pb+Pb collisions with a muon 
spectrometer, covering the forward rapidity region($2.5<y<4$). The 
analysis of the inclusive J/$\psi$ production in the first Pb+Pb data 
collected in the fall 2010 at a center of mass energy of 
$\sqrt{s_{\rm NN}}=2.76$~TeV is discussed. 
Preliminary results on the nuclear modification factor ($R_{AA}$) and 
the central to peripheral nuclear modification factor ($R_{CP}$) are presented.
\end{abstract}

\keywords{Quark-Gluon Plasma, Muon Spectrometer,Quarkonia, LHC, Heavy Ion Collisions}

\maketitle

\section{Introduction}

The quarkonia produced in ultra-relativistic heavy-ion collisions are considered as excellent 
effective probes of the strongly interacting medium and also proposed as a signature of 
deconfinement~\cite{Satz86}. In such hot deconfined matter, the color screening 
dissolves the binding of the heavy quark anti-quark pair. 
This leads to the higher excited quarkonium states 
to dissolve at lower temperature. Studying the suppression pattern of quarkonia in 
AA collisions (where the plasma of deconfined quarks and gluons, 
the Quark-Gluon Plasma (QGP) is expected to form), along with the 
comparative quarkonia production in pp collisions will allow one 
to set an estimate of the initial temperature of the medium~\cite{Satz86}. 
Extensive experimental results at SPS~\cite{NA60}(including feeddown from 
other less bound resonances like $\psi^{'}$  and $\chi_{c}$) and RHIC~\cite{Kikola:2009us} of  
J/$\psi$ production in AA collisions clearly indicate that even the 
strong bound J/$\psi$ ground state is suppressed.

However at Large Hadron Collider(LHC) energies the J/$\psi$ 
production could be even enhanced due to the coalescence 
of un-correlated $c\bar{c}$ pairs in the medium which can cause a regeneration~\cite{BraunMunzinger:2000px}.
Initial state effects like  modifications of the parton distribution functions in 
the nucleus relative to the nucleon (also known as shadowing) need to be taken in account~\cite{Satz86}. 
The final state effects like nuclear absorption are expected 
to be practically irrelevant at LHC energies. Studying the pA collisions 
at LHC energies is henceforth crucial to quantify the role of initial shadowing effects.
Using the Muon Spectrometer~\cite{Aamodt:2011gj} charm and beauty particles 
can be measured in the forward rapidity region via its di-muon decay. The ALICE Muon Spectrometer 
physics program~\cite{Aamodt:2011gj} is based on the  measurement of 
heavy-flavor production in forward rapidity region ($2.5<y<4$) for pp, pA and AA collisions at LHC energies.

\begin{figure}[htbp]
\includegraphics[width=8cm]{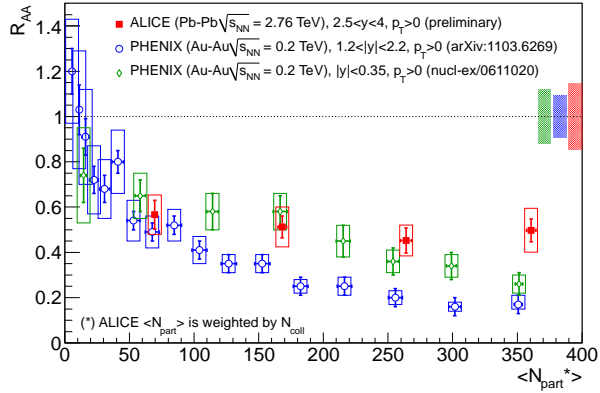}
\caption{\label{fig:RAA} J/$\psi$ $R_{\rm AA}$ in Pb+Pb at $\sqrt{s_{\rm NN}}=2.76$~TeV
as a function of $\langle N_{\rm part} \rangle$ compared with PHENIX results in Au+Au collisions at $\sqrt{s_{\rm NN}}=200$~GeV.}
\end{figure}

\begin{figure}[htbp]
\includegraphics[width=8cm]{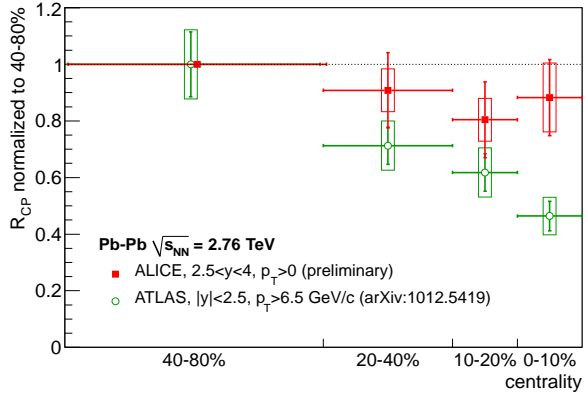}
\caption{\label{fig:RCP} J/$\psi$ $R_{\rm CP}$ as a function of centrality compared with ATLAS 
results. Error bars represent the statistical uncertainties, open boxes show the 
centrality-dependent systematic uncertainties while the centrality independent uncertainties are represented 
by filled boxes.}
\end{figure}

\section{$R_{\rm AA}$ and $R_{\rm CP}$}

The first Pb+Pb collisions were delivered by LHC at a centre of mass 
energy $\sqrt{s_{\rm NN}}=2.76$~TeV in the fall of 2010 at an un-explored regime 
of almost 14 times higher than top RHIC energy. The results presented here are 
based on sample corresponding to an integrated luminosity of 2.7 $\mu$b$^{-1}$.
The nuclear modification factors defined as $R_{\rm AA}$ or $R_{\rm CP}$ 
allow us to quantify the medium effects on J/$\psi$ production. 
$R_{\rm AA}$ gives the deviation in  J/$\psi$ yields
from AA collisions relative to the scaled 
(according to the number of binary nucleon-nucleon collisions) 
yields of  J/$\psi$ from pp collisions. 
Figure~\ref{fig:RAA} shows the  inclusive J/$\psi$ $R_{\rm AA}$ 
measured at forward rapidity($2.5<y<4$) as a function of  
the average number of nucleons participating to the 
collision ($\langle N_{\rm part} \rangle$) which has been 
calculated using the Glauber model. $\langle N_{\rm part} \rangle$ has been weighted by the number of 
binary nucleon-nucleon collisions ($N_{\rm coll}$) 
due to the bias caused by large centrality bins. This correction is small, 
except for the most peripheral bin where $\langle N_{\rm part} \rangle$~=~46 
while the weighted value is 70. These results show weak centrality dependence 
and an integrated $R_{\rm AA}^{0-80\%}=0.49\pm0.03({\rm stat.})\pm0.11({\rm syst.})$. 
Comparison with the RHIC measurements at $\sqrt{s_{\rm NN}}=200$~GeV~\cite{PHENIX07} 
from the PHENIX experiment shows that the 
inclusive J/$\psi$ $R_{\rm AA}$ at 2.76~TeV in the ALICE forward rapidity 
region are higher than that measured at 200~GeV in the rapidity domain of $1.2 < |y| < 2.2$. 
However, the midrapidity values at 200~GeV (except in the 
most central collisions) are closer. The contribution from the B feed down to the J/$\psi$ 
production in our rapidity and $p_{\rm T}$ domain has been measured and estimated to be $\approx10\%$ 
in pp collisions at $\sqrt{s_{\rm NN}}=7$~TeV~\cite{LHCb}. 
$R_{CP}$ can provide similar information based on the relative yield in central(C) and peripheral(P)
collisions scaled by the mean number of binary collisions, but does not depend on the
reference pp system. Figure~\ref{fig:RCP} shows the ALICE forward rapidity measurements of  J/$\psi$ $R_{\rm CP}$ 
results compared with the ATLAS measurements for the same centrality classes~\cite{ATLAS}.
The J/$\psi$ mesons measured at forward rapidity down to $p_{\rm T}=0$ 
are less suppressed than the high-$p_{\rm T}$ J/$\psi$ mesons at midrapidity (80\% of 
the J/$\psi$ particles measured by ATLAS have a $p_{\rm T}$ larger than 6.5 GeV/$c$).

\section{Summary and Outlook}

Nuclear modification factors $R_{\rm AA}$ or $R_{\rm CP}$ are measured in the first heavy-ion 
2010 run of LHC. The inclusive J/$\psi$ production in Pb+Pb collisions at $\sqrt{s_{\rm NN}}=2.76$~TeV
shows a notable suppression and no significant dependence on centrality.
At central events, a smaller suppression has been found compared to PHENIX results at RHIC energies.
The regeneration mechanism may explain these experimental results.
The roles of the suppression and regeneration mechanisms could be 
disentangled by quantifying the initial shadowing effect with data 
from the forthcoming p+Pb run at LHC, planned for late 2012.


\end{document}